\documentclass[aps,twocolumn,pra,showpacs,floatfix]{revtex4}
\usepackage{epsfig}
\usepackage{graphicx}
\usepackage{dcolumn}
\usepackage{amssymb,amsmath}
\usepackage{mathrsfs}
\begin{document}

\newcommand{\Z}{Z_{\rm eff}}
\newcommand{\Zr}{\Z^{(\rm res)}}
\newcommand{\eps}{\varepsilon}
\newcommand{\eb}{\varepsilon _b}

\title{Effect of atomic electric quadrupole moment on positron binding}

\author{C. Harabati, V. A. Dzuba, and V. V. Flambaum}

\affiliation{School of Physics, University of New South Wales, Sydney 2052,
Australia}

\begin{abstract}
Effect of the electric quadrupole moment, $Q$, is studied for
positron-atom bound systems. It is demonstrated that for $Q >50$ a.u.
the electric quadrupole potential is sufficiently strong to
bind positron (or electron) even in the absence of the dipole polarization potential.  Such large values of $Q$ are not known  for atomic ground states, however, they exist in molecules and excited atoms.  
In the state $2s2p~^3P^o_2$ of beryllium, the quadrupole contribution makes difference between stable bound state and decay to Be$^+$ ion and positronium. In a majority of atoms the quadrupole contribution is small and can be neglected.
\end{abstract}


\maketitle
\section{Introduction}

Unperturbed central potential of an atom is positive on all distances and can not bind positron. 
Only when correlations between positron and atomic electrons are
taken into account, positron can be bound to a majority of
atoms~\cite{HDF14} (see also \cite{MBR02,DFGH12,RM97,SC98,MZB08,BM10,DFGH99,DFH00,DFGK95} and references therein). A simple explanation of the positron binding is 
polarization of the atom by positron field. Electron-positron
attraction shifts the electron cloud towards the positron producing
electric dipole moment of the atom. This induced electric dipole
moment creates polarization potential which on large distances
behaves like $-\alpha/2r^4$, where $\alpha$ is static dipole
polarizability of the atom. 
 However, many  atoms  also have a static quadrupole moment
which produces a long-range potential which decays slower ($\sim 1/r^3$) than the polarization potential ($\sim 1/r^4$).
Therefore, it looks interesting to investigate the role of the quadrupole potential
in the positron binding.

There are many different techniques used to calculate positron binding
energy to atoms. The variational
 and configuration interaction calculations may easily  include the quadrupole contribution. In other
calculations based on the  correlation potential~\cite{DFGK95,DFGH99,DFH00} and coupled-cluster
single-double~\cite{DFGH12,HDF14} approaches the quadrupole potential contribution was not included since the positron was assumed to be in $s$-wave. Note, that
assuming positron being in $s$ state does not mean that contribution
of  higher angular momenta are totally neglected. For example, in
the case of correlation potential method~\cite{DFGK95,DFGH99,DFH00} binding energy is
related to the expectation value of the correlation potential $\hat
\Sigma$: $\epsilon \sim \langle s |\hat \Sigma | s \rangle$, where $s$
is the positron wave function and higher values of both electron and
positron angular momenta are included into the perturbation theory calculation of $\hat
\Sigma$. Therefore, the effect of the virtual positron formation is
taken into account (see, e.g. \cite{DFGK95,DFGH99,DFKMS93}). 

In spite of the fact that many calculations do include the quadrupole
contribution, it was never presented separately. However, it is useful
to know it to judge whether it can be at least partly responsible for
the difference in results in different calculations. It is also
important to know whether that contribution can be large enough to
provide the difference between binding and not binding. In this paper
we study the quadrupole contribution using simple variational approach.

\section{Variational energy of a positron in an atom }

Positron trial wave function in the presence of electric dipole and
quadrupole potentials is taken in the form
\begin{align}
\Psi(r,\theta)=A(r-a)e^{-\kappa r}(\sqrt{1-\beta^2}Y_{00}+\beta
Y_{20}) \label{eq:psi}  
\end{align}
where $A$ is the normalization constant, $Y_{lm}(\theta,\phi)$ are 
spherical harmonics, $\kappa$ and $\beta$ are variational parameters,
and $a$ is a cut-off distance from the nucleus. 
It is assumed that the
wave function is zero at $r<a$. 
The potential energy of the positron for $r > a$ is
\begin{equation}
V(r,\theta) = -\frac{e^2\alpha}{2r^4}+\frac{eQ}{2}\frac{P_2(\cos\theta) }{r^3}
\label{eq:potential}
\end{equation}
where $\alpha$ is the polarizability of the atom, $Q$ is its
quadrupole moment, and $P_2(\cos\theta)=(3 \cos^2\theta-1)/2$ is
Legendre polynomial. There is also additional infinite potential
barrier at $r=a$ which simulates the effect of positron repulsion from
the positive electrostatic potential inside the atom. The mixing of $s$ and $d$ waves provides localisation of the positron wave function in equatorial or polar areas depending on
 the sign of $Q$.

Using Schr\"{o}dinger equation positron energy can be written as
\begin{equation}
E(\kappa,\beta)= \frac{\int d^3r \left( \frac{\hslash^2}
    {2m}\nabla\Psi^\ast \cdot \nabla\Psi+\Psi^\ast V\Psi \right)
}{\int d^3r\Psi^\ast\Psi } 
\label{eq:variation}
\end{equation}
Substituting (\ref{eq:psi}) into (\ref{eq:variation}) leads to explicit
form of $E(\kappa,\beta)$: 
\begin{equation}
\frac{4ma^2}{\hslash^2}E(\kappa,\beta)=\mathscr{E}(x,\beta)=
\frac{x^2f(x,\beta)}{x^2+6x+12} 
\label{eq:energy}
\end{equation}
where $x=2a\kappa$ is a dimensionless variational parameter and
$\mathscr{E}(x,\beta)$ is the dimensionless variational energy. The
function $f(x,\beta)$ can be written in the form  
\begin{align}
f(x,\beta)=&-\Omega x^3+(1/2-\Omega-\Lambda\xi)x^2+\nonumber \\
&+(\Lambda\xi+1)x+12\beta^2+2+\nonumber\\
&+(\Omega x^3+(2\Omega+\Lambda\xi)x^2)xe^xE_1(x)
\label{eq:f} 
\end{align} 
where 
\begin{equation}
\Omega=\frac{me^2\alpha }{\hslash^2 a^2},
\label{eq:omega}
\end{equation}

\begin{equation}
\Lambda=2\frac{me Q}{\hslash^2 a},
\label{eq:lamda}
\end{equation}

\begin{equation}
\xi=\beta\sqrt{\frac{1-\beta^2}{5}}+\frac{\beta^2}{3},
\label{eq:xi}
\end{equation}


and 
\begin{equation}
E_1(x)=\int_x^\infty dt\frac{e^{-t}}{t}
\label{eq:expint}
\end{equation}
is the exponential integral \cite{AS64}. 

Variation of the energy (\ref{eq:variation}) with respect of the
parameters $x$ and $\beta$ requires that $$ \frac{\partial
  \mathscr{E}(x,\beta)}{ \partial x}=0,~~ \frac{\partial
  \mathscr{E}(x,\beta)}{\partial \beta}=0.$$
Solving these equations for $x$ and $\beta$ would lead to the ground
state energy $ \mathscr{E}( x_0, \beta_0)$ of the positron in an atom
specified by three parameters, the cut-off parameter $a$, the polarizability 
$\alpha$, and quadrupole moment $Q$. 

Partial derivative of the energy (\ref{eq:energy}) with respect to
$\beta$ leads to  
\begin{equation}
\frac{\partial \mathscr{E}}{\partial
  \beta}=\Lambda\frac{\partial\xi}{\partial
  \beta}((xe^xE_1(x)-1)x^2+x)+24\beta=0. 
\label{eq:second}
\end{equation}
This equation can be used to express $\beta$ in terms of the other
parameter $x$, 
\begin{equation}
\beta=\pm \sqrt{(1\pm \lambda(x))/2}, 
\label{eq:beta}
\end{equation} 
where $\lambda (x)$ is a simple function of $x$: 
\begin{equation}
\lambda(x)=\sqrt{1-\frac{1}{1+5
    \left(\frac{\hslash^2}{me}\frac{a}{Q}\frac{6}{g(x)}+\frac{1}{3}
    \right)^2 }}. 
\label{eq:lambda}
\end{equation} 
The function $g(x)= (xe^xE_1(x)-1)x^2+x$ is the same as the
$x$-dependent part of the equation (\ref{eq:second}). It is positive for all
$x > 0$ and $g(x)\sim 2$ asymptotically at $x\rightarrow \infty$. 

We substitute the expression (\ref{eq:beta}) for $\beta$ into 
second variational equation $\frac{\partial \mathscr{E}}{\partial
  x}(x,\beta(x))=0$. For each root $x_0$ of this equation there are
four different values of $\beta$ found from Eq. (\ref{eq:beta}).
 
Final equation to solve for the extrema of the energy,
which depends only on $x$ after substituting $\beta(x)$, has the form
\begin{widetext}
\begin{align}
\frac{-(\Omega+1/2) x^4+(1-4\Omega-5\Lambda\xi)x^3+(1-12\beta^2+3\Lambda\xi)x^2+4(1+
3\beta^2)x+(6\Omega x^4+5(2\Omega+\Lambda\xi)x^3)xe^xE_1(x)}{-\Omega x^5+(1/2-\Omega-\Lambda\xi)x^4+(1+\Lambda\xi)x^3+2(1+
6\beta^2)x^2+(\Omega x^5+(2\Omega+\Lambda\xi)x^4)xe^xE_1(x)}=\nonumber\\
-\frac{x^2+4x+6}{x^2+6x+12}
\label{eq:first} 
\end{align}
\end{widetext} 
This equation is to be solved numerically for the roots $x=x_0$ if they exist.  

\section{Results and discussion}\label{sec:result}
\begin{table*}
\caption{Energy shift, $\Delta E$ of a bound positron in an atom due to the electric quadrupole moment $Q$ of the atom. $I$ is the ionization energy and $\alpha$ is the static dipole polarizability of the atomic state. The parameter $a$ included in the positron trial function shows the minimal  distance  between  the positron and the nucleus. $E$ is the positron energy  without quadrupole contribution except Be atom case where the quadrupole effect has already been included in the positron energy, \cite{BP13}}
\label{t:1}
\begin{ruledtabular}
\begin{tabular}{rll rrrrr rrrr}
\multicolumn{2}{c}{} & 
\multicolumn{1}{c}{} &
\multicolumn{1}{c}{$I\footnotemark[3]$} &
\multicolumn{1}{c}{$\alpha$\footnotemark[1]} &
\multicolumn{1}{c}{a}&
\multicolumn{1}{c}{Q} &
\multicolumn{1}{c}{$E\footnotemark[2]$}&
\multicolumn{1}{c}{$\Delta E$}\\
\multicolumn{1}{c}{$Z$} &Atom &
\multicolumn{1}{c}{state} &
\multicolumn{1}{c}{(eV)} &
\multicolumn{1}{c}{(a.u.)} &
\multicolumn{1}{c}{(a.u.)} &
\multicolumn{1}{c}{(a.u.)} & 
\multicolumn{1}{c}{(meV)} &
\multicolumn{1}{c}{(meV)}\\
\hline
\multicolumn{9}{c}{Ground states}\\ 
66 & Dy & $4f^{10}6s^2$ $^5I_8$  & 5.939 & 162.02 & 2.5878 & 0.0234 & -1438 & -5.36$\times 10^{-4}$\\
68 & Er & $4f^{12}6s^2$ $^3H_6$  & 6.107 & 150.12 & 2.530  & 0.0139 & -1346 & -1.98$\times 10^{-4}$\\
77 & Ir & $5d^76s^2$ $^4F_{9/2}$ & 8.967 & 50.26  & 1.8606 & 0.75 & -101  & -0.6549 \\ 
\hline
\multicolumn{9}{c}{Excited states}\\
4 &  Be & $2s2p$ $^3P^o_2$      &  6.597 & 38.33 &    1.60774      & 4.28\footnotemark[6],4.53\footnotemark[7]& -236\footnotemark[8] & -41.98\\
13 & Al & $3s^23p$ $^2P^o_{3/2}$ & 5.972 & 44.97 & 1.81304 & 5.6, 5.06\footnotemark[4] & -0 & -28.67 \\  
49 & In & $5s^25p$ $^2P^o_{3/2}$ & 5.512 & 67.45  & 2.13042 & 5.88\footnotemark[5] & -114 & -23.15 \\    
\end{tabular}
\footnotetext[1]{Ground-state atomic static dipole polarizabilities from Ref. \cite{Miller}}.
\footnotetext[2]{Recommended positron energies in Ref. \cite{HDF14}}.
\footnotetext[3]{Ionization potential from NIST atomic database \cite{NIST}}.
\footnotetext[4]{Experimental value from Ref.\cite{ASW67}}.
\footnotetext[5]{Experimental value from Ref.\cite{SO93}}.
\footnotetext[6]{EQM of Be atom from Ref.\cite{CB91}}.
\footnotetext[7]{EQM of Be atom from Ref.\cite{SO93}}.
\footnotetext[8]{Positron energy in the excited state of Be from Ref.\cite{BP13}}.
\end{ruledtabular}

\end{table*}

To use Eq.(\ref{eq:first}) for calculation of the positron
energy we need to know atomic polarizability $\alpha$, atomic
quadrupole moment $Q$ and the value of the cut-off parameter $a$ of the trial positron wave function (\ref{eq:psi}). For
polarizabilities and quadrupole moments we use the values which can be
found in the literature (either experimental or theoretical) and we
treat the cut-off parameter $a$ as a fitting parameter. We choose its
value to fit the most accurate calculations of the positron energy. 
The value of the quadrupole contribution is found as a
difference between the energy at given value of the quadrupole
moment and the value found at $Q=0$.   

In cases when atomic polarizability or quadrupole moment cannot be
found in the literature we calculate them using the configuration
interaction (CI) technique~\cite{DFK96,DF08}.
Static scalar polarizability of an atom in a state $\gamma$ is given by
\begin{equation}
\alpha_{\gamma} = \frac{2}{3(2J_{\gamma}+1)} \sum_n \frac{\langle
  \gamma||\mathbf{D}|| n   \rangle^2}{E_{\gamma} - E_n},
\label{alpha0} 
\end{equation}
while electric quadrupole moment is given by
\begin{equation}
Q_{\gamma}=-2 \sqrt{
  \frac{J_{\gamma}(2J_{\gamma}-1)}{(2J_{\gamma}+3)(2J_{\gamma}+1)(J_{\gamma}+1)}
}\langle\gamma \Vert r^2\Vert\gamma \rangle \label{eq:Q}.
\end{equation}
Here $|\gamma \rangle$ and $| n \rangle$ are many-electron states
found in the CI calculations.

About half of all atoms do not have quadrupole moments in the ground state due to
small value of the total angular momentum $J$ (one needs $J \ge
1$). Atoms with open $d$ or
$f$ shells have large angular momenta. In Table \ref{t:1} we present three such examples, Dy, Er
and Ir atoms. Polarizabilities $\alpha$ are taken from Ref.~\cite{Miller},
quadrupole moments $Q$ are calculated using the CI technique, the
cut-off parameter $a$ is chosen to fit calculated positron energies
presented in Ref.~\cite{HDF14}. Since the calculations in \cite{HDF14}
are done under assumption that the positron is in $s$ wave and
therefore cannot interact with the atomic quadrupole moment the
fitting is done for $Q=0$. After the cut-off parameter $a$ is found,
Eq. (\ref{eq:first}) with values of $Q$ from the table is used to
calculate new bound energies. The resulting energy shift $\Delta E$ is
the quadrupole contribution to the energy. We see that it is
small in all three cases. This justifies neglecting the quadrupole
contribution in calculations of Ref.~\cite{HDF14}.

Small value of the quadrupole contribution means that it can be
treated by means of the perturbation theory. The first-order contribution
for positron in $s$ wave is zero, therefore expansion starts from the second
order and quadrupole contribution is proportional to the square of the
quadrupole moment. Using Ir atom as a reference point we can estimate the quadrupole contribution to the positron energy level for
any atom with a small quadrupole moment:
\begin{equation}\label{eq:shift}
\Delta E=-0.6549\left(\frac{Q}{0.75} \right)^2 \,\,{\rm meV}, 
\end{equation}
where 
$Q$ is in atomic units ($ea_0^2$). The
values given by this formula differ from those presented in
Table~\ref{t:1} by 13\% for Er and 19\% for Dy. We expect similar
accuracy for other atoms with small quadrupole moments.

In Table~\ref{t:1} we also present three other results for the quadrupole
contribution to the positron energy. We consider excited state of
Be atom for which accurate calculations of the positron energy
is available~\cite{BP13} and we consider upper ($^2P^o_{3/2}$) components of the fine
structure doublets of the ground $p$ state of Al and In. In all these cases the value of the quadrupole moment is relatively
large. So is the quadrupole shift. The case of Be atom is interesting
because the quadrupole contribution makes important difference for the
positron energy. The calculated energy \cite{BP13}(-236 meV) relative to the atom plus free positron does include the quadrupole contribution. However, if this contribution is neglected, the system becomes unstable against emission of positronium (Be+e$^+ \ \rightarrow$ Be$^+$ + Ps).

\begin{figure*}
\centering
\epsfig{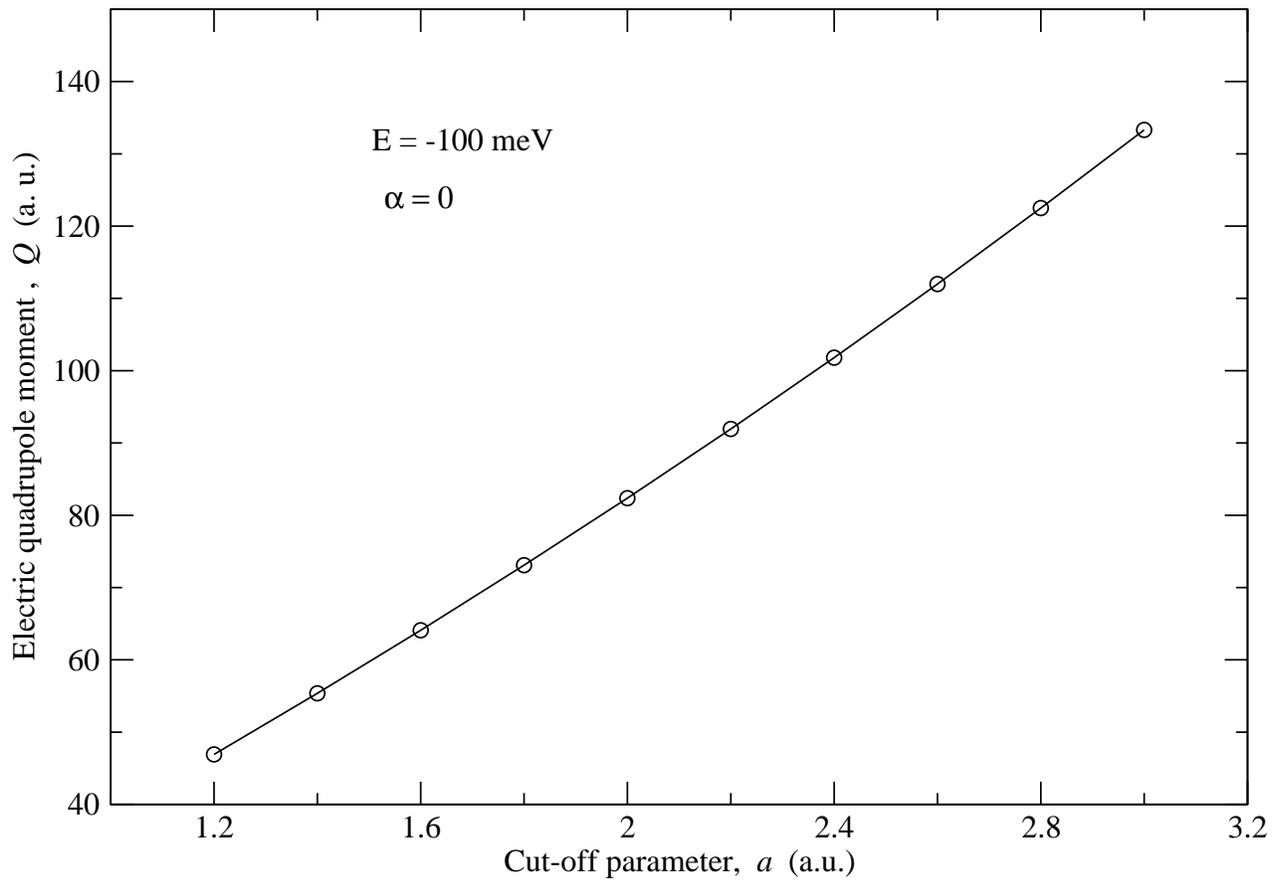}
\caption{The required electric quadrupole moment $Q$ versus the the small distance cut-off parameter of the positron wave function, $a$, in a.u., to get a bound state with the energy $E= -100$ meV. The dipole polarizability is assumed to be zero, $\alpha=0$.} 
\label{f:1}
\end{figure*}

   In some excited atomic states the polarizability may be very small or even negative \cite{MSC10}.
Therefore, it is interesting to check if the quadrupole alone (for $\alpha=0$) may provide the positron binding. 
 We use
Eq.~(\ref{eq:first}) to estimate what value of the quadrupole moment is
needed to provide the positron binding. Fig.~\ref{f:1} shows a plot of the quadrupole
moment corresponding to the $-100$ meV positron energy as a function of the
cut-off parameter $a$. We use an estimation $a \sim 1/I$ to find
reasonable range of values for $a$. Here $I$ is ionization potential,
and both values $a$ and $I$ are in atomic units. We see that required
values of the quadrupole moments are large. No atom in the ground
state has so large quadrupole moment.  
However, quadrupole moment can be large in excited state, $Q\sim
\nu^4$, where $\nu$ is the effective principle quantum number
($E=-1/(2\nu^2)$).    Large values of $Q$ proportional to their squared size may also exist in molecules.  

\section{Conclusion}

A simple quantum mechanical variational method is used to estimate the
contribution of the positron interaction with atomic quadrupole moment
to the positron energy in an atom. It is found that
the contribution is small for the binding in the ground state. This
validates the calculations in which this contribution is neglected. The quadrupole contribution can be significant in excited
states. For example, in excited state of beryllium it makes a difference
between a bound state and decay into beryllium positive ion and
positronium. 

Quadrupole contribution can probably play important role in positron
binding to molecules which have large quadrupole moments. 

\begin{acknowledgements}
This work was funded by the Australian Research Council.  
\end{acknowledgements}

\end{document}